\documentclass[final,1p,times]{elsarticle}

\usepackage{subfigure}
  
\usepackage{amssymb}

\journal{Nuclear Physics A}

\begin{document}

\begin{frontmatter}


\title{Jet-Medium Interactions with Identified Particles}
\author{Anne M. Sickles}
\ead{anne@bnl.gov}
\address{Brookhaven National Laboratory, Upton NY 11973\fnref{label3}}





\begin{abstract}
Identified particles have long been of great interest at 
RHIC in large part because of the baryon/meson differences 
observed at intermediate $p_T$ and the implications for hadronization
via quark coalescence.  With recent high statistics data identified
particles are also now central to understanding the details of
the jet-medium interactions and energy loss and hadron formation at intermediate
and high $p_T$.  In particular, high $p_T$ identified particle spectra along with
two-particle correlations triggered with direct
photons, neutral pions or electrons from heavy flavor decay with
hadrons can provide information about how 
medium modifications to jet fragmentation depend on parton type.
I will review recent results with identified
particles both in heavy ion systems and the reference 
measurements in p+p collisions.
\end{abstract}




\end{frontmatter}


\section{Hard Scattering at RHIC}\label{intro}
Hard probes of relativistic heavy ion collisions have long been 
a valuable probe of the hot matter created in relativistic heavy ion
collisions.  Their main value comes from the fact that their production
is both calculable in perturbative QCD and measurable in proton-proton collisions, 
providing quantified expectations for heavy ion collisions.  Deviations from proton-proton
expectations provide measurements of the effects of the hot dense matter on
the propagation of fast partons.
In high energy experiments hard probes are often measured through
directly reconstructing full jets.  Such measurements are not 
naturally suited to the large soft background of heavy ion collisions,
especially when one wants to study the modification of jets from baseline proton-proton collisions
measurements (for a discussion of jet reconstruction in heavy ion collisions 
see~\cite{sevil_qm09,lai_qm09}).
More experimentally straightforward are single particle spectra and two particle correlations 
which provide complementary
observables with relatively large rates.  Particle identification
provides a valuable experimental handle for changing the probe partons (e.g.
light quarks, gluons, heavy quarks, photons) and for studying the effect
of the matter on hadron formation~\cite{friesprc,hwa1,stananne,vitev}.  Current
results from RHIC allow detailed quantitative studies of how jet modifications
depend on parton type and provide insights on the interactions of the matter
with fast partons and the hadron formation process.
 
\section{Single Particle Production}
\begin{figure}[t]
\centering
\includegraphics[width=0.9\textwidth]{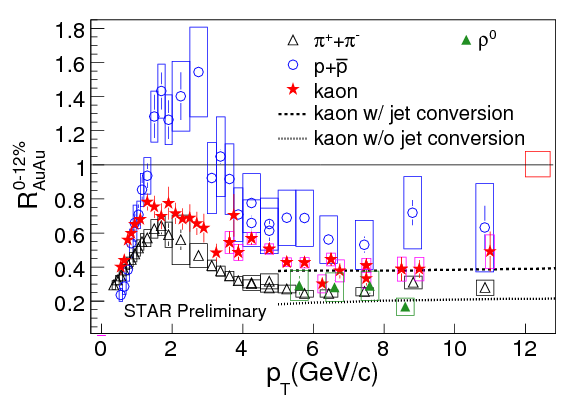}
\caption{(color online) Nuclear modification factor, $R_{AA}$ for charged $\pi$, $K$ and $p$+$\bar{p}$
for central collisions~\protect\cite{yichunqm09}.   }
\label{raa}
\end{figure}

The most stringent constraint on the opacity of the hot matter, within a particular
theoretical model, comes from $\pi^0$ $R_{AA}$~\cite{ppg079}.  The $R_{AA}$ of other particles
can provide additional information about parton-medium interactions and hadron formation.
Expectations for the $R_{AA}$ of protons and anti-protons were that they would be more
suppressed than $\pi^0$ since gluon jets were thought to produce more baryons than quark jets do
(however, recent fragmentation functions incorporating the STAR $p,\bar{p}$ spectra
cast some doubt on this~\cite{dss07}).
Since gluons have a larger QCD color factor, they should lose more energy.  However,
recent measurements from STAR show that $R_{AA}(p,\bar{p}) > R_{AA}(\pi^0)$, 
Fig.~\ref{raa}.  At intermediate $p_T$, $2<p_T<6$~GeV/c, this difference is widely
thought to be from recombination (see Ref~\cite{reco_review} and references therein), 
though realistic calculations taking into account
spatial correlations, energy and momentum conservation and gluons have not yet been done.

For $p_T>6$GeV/c, the difference between $R_{AA}(p,\bar{p})$ and
$R_{AA}(\pi^0)$ is smaller but still significant and is not understood.  One
possible idea is that as the jet partons traverse the matter they change flavor
after scattering on matter partons~\cite{conversion_ko,conversion_fries},
for example a fast quark could scatter
off a medium gluon and emerge as a fast gluon.  Discovery of these conversions 
would be extremely interesting as it would provide a means to study the 
mean free path of partons in the matter. However, 
this mechanism could make $R_{AA}(p,\bar{p})
\approx R_{AA}(\pi^0)$, but cannot make $R_{AA}(p,\bar{p})>R_{AA}(\pi^0)$ so
it cannot, in itself, explain the current data. 

Alternatively, it has been proposed that baryons and anti-baryons observed
at RHIC at high $p_T$ are produced in higher twist QCD processes in the initial
state~\cite{stananne}.  Processes such as $uu\to p\bar{d}$ can occur in QCD, but
they are typically rare compared to production via jet fragmentation.  However,
since the protons produced directly in the hard scattering are small enough to 
propagate through the matter without interacting, similar to a direct photon.  
Such $p$ and $\bar{p}$ would also be produced in proton-proton collisions, however
the dense matter in the heavy ion collision selects out these protons because the partons that would lead to 
fragmentation $p$ and $\bar{p}$ production lose energy.
This mechanism naturally leads to $R_{AA}(p,\bar{p}) > R_{AA}(\pi^0)$.  The decreasing
$v_2$ of $p$ and $\bar{p}$~\cite{shengli} observed by PHENIX is also easily explained
if an increasing fraction of the $p$ and $\bar{p}$ is insensitive to the path
length through the matter and this also explains the centrality dependence
of $p$ and $\bar{p}$ triggered correlations at intermediate $p_T$\cite{ppg072}. 

\section{Two Particle Correlations}
Two particle correlations have been used extensively to investigate
jet production at RHIC.  They are complementary to single particle
observables,  especially in that they are expected to have a different
sensitivity to geometry.  Single particle observables are thought to be strongly biased
toward particles toward partons which have lost less than the
average amount of energy which is dominated by those with a short path length through the matter.  Requiring a correlated particle
changes the surface bias depending on the $p_T$ and $\Delta\phi$ between the
trigger and associated particle. At high $p_T$, two-particle back-to-back correlations can
help constrain the path length dependence of energy loss.  Previous
measurements with charged hadrons have shown a strong suppression of
the away side hadrons when a high $p_T$ trigger is required~\cite{stardijet,ppg083}.
At intermediate $p_T$ with hadron-hadron correlations a third peak at $|\Delta\phi-\pi|\approx$1~radian, 
termed {\it the
shoulder} is observed~\cite{ppg083,ppg032,starlowpt}.  Both the trigger particle
and $p_T$ dependence of this structure are of great interest in determining
its origin and sensitivity to properties of the matter.
New measurements with the 2007 RHIC run allow further exploration of
this with identified triggers and smaller $p_T$ bins.  Identified trigger
particles are important because of the dependence of $R_{AA}$ on the
hadron species out to the highest measured $p_T$.

\subsection{Direct Photon-Hadron \& $\pi^0$-Hadron Correlations}
Direct photon-hadron ($\gamma_{dir}$-h) correlations have long been considered an important
observable in studying energy loss.  The photons are primarily produced
in hard scattering processes in the initial state.  Photons
do not lose energy in the matter and therefore, at leading order, 
provides a measure of the away side jet energy (initial state $k_T$ and NLO
effects modify this somewhat, see Ref.~\cite{frantzhp08} for recent measurements).  Because
the photons do not lose energy 
they are not surfaced biased as hadron triggered measurements
are.  

\begin{figure}
\includegraphics[width=0.55\textwidth]{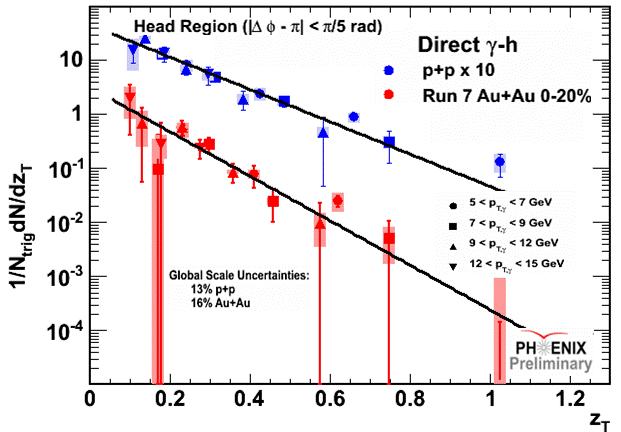}
\includegraphics[width=0.40\textwidth]{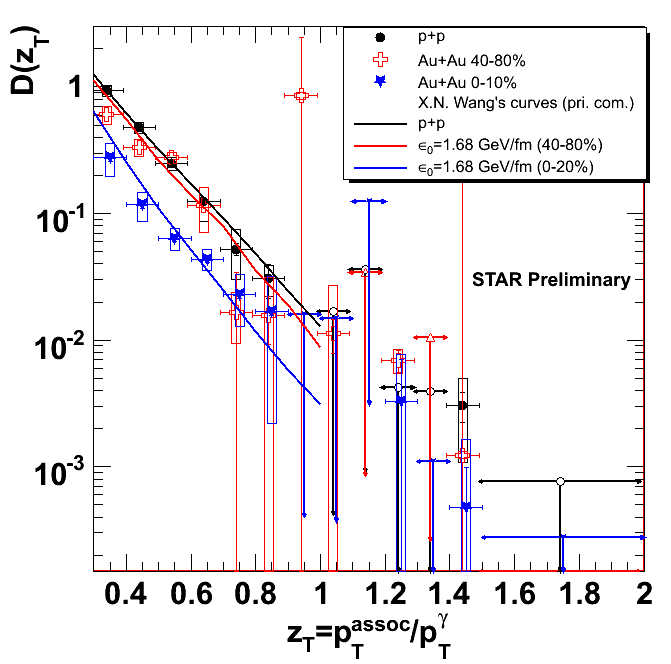}
\caption{(color online) $z_T$ distributions for PHENIX~\protect\cite{megan_qm09} (left) and STAR~\protect\cite{hamed_qm09}
 (right) $\gamma_{dir}$-h correlations for Au+Au and p+p collisions.}
\label{ztdist}
\end{figure}

However, the measurements are extremely difficult because of the large background
from photons from meson (primarily $\pi^0$) decay: $\gamma_{decay}$.  
What is measured is the weighted average of the conditional yield associated
with $\gamma_{dir}$-h and $\gamma_{decay}$-h:
\begin{equation}
Y_{\gamma_{incl}-h} = \frac{N_{direct}}{N_{incl}} Y_{\gamma_{direct}-h}
+ \frac{N_{decay}}{N_{incl}} Y_{\gamma_{decay}-h}
\end{equation}
The two unknowns are then the fraction of the photons which are direct and the
 $\gamma_{decay}$-h conditional yield.   The PHENIX and STAR methods for
determining these quantities are quite different and are explained 
elsewhere~\cite{ppg090,megan_qm09,hamed_qm09}.
PHENIX
recently published results from the 2004 RHIC run which show first 
measurements of $\gamma_{dir}$-h correlations in Au+Au and p+p collisions~\cite{ppg090}.
Preliminary results from the 2007 RHIC run with higher statistics 
are available from both PHENIX and STAR.  
Since the $\gamma_{dir}$ provides a measurement of the jet energy, the conditional
yield as a function of $z_T \equiv p_{T,h}/p_{T,trig}$ is approximately
the jet fragmentation function into hadrons. At leading order
the away side jet is expected to be a (anti-)quark jet because $\gamma_{dir}$
production is largely via $qg \to q\gamma$.  The data are shown in Figure~\ref{ztdist}. The exponential 
slope of the quark fragmentation function is $\approx$8~\cite{ppg029}
and a fit to the p+p results gives a slope of 6.89$\pm$0.64 (consistent with
quark jet fragmentation functions~\cite{ppg029}).
Fitting the PHENIX Au+Au data gives a slightly softer slope: 9.5$\pm$1.4.

It is useful to compare $\gamma_{dir}$-h and $\pi^0$-h correlations.
$\pi^0$-h correlations show a nearly constant suppression with $p_{T,trig}$
and $p_{T,h}$ for a wide $p_T$ range in central collisions. 
In contrast $\pi^0$-h correlations are biased toward small medium path lengths and
the $\pi^0$ does not carry the full jet energy. 
Therefore, differences between away side yields of $\gamma_{dir}$-h and $\pi^0$-h provide information
about the geometry and the energy dependences of energy loss.
Within the current uncertainties the data are consistent between
these two channels.  With the present uncertainties, theoretical predictions
do not predict a measurable difference between these channels in the $p_T$ ranges
measured~\cite{zoww_gamma,zoww_hadron,renk_gamma,renk_hadron}.  

\begin{figure}
\includegraphics[width=0.55\textwidth]{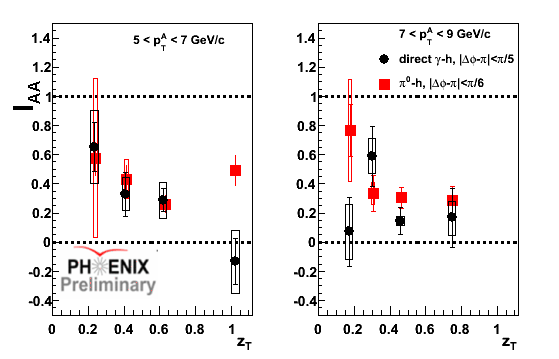}
\includegraphics[width=0.40\textwidth]{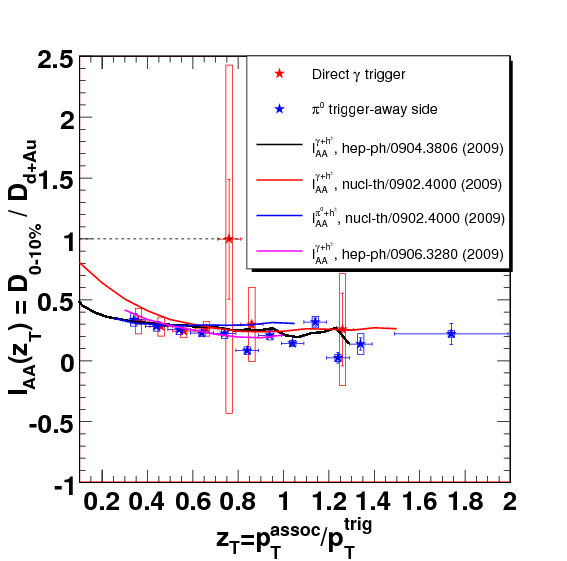}
\caption{(color online) 
$I_{AA}$ for $\gamma_{dir}$-h and $\pi^0$-h away side yields from PHENIX\protect\cite{megan_qm09} (left)
and STAR~\protect\cite{hamed_qm09} (right).  Both
results show no $p_T$ dependence or any significant difference between the suppression of $\pi^0$
and $\gamma_{dir}$ triggers.}
\label{iaa_comp}
\end{figure}

\begin{figure}
\includegraphics[width=\textwidth]{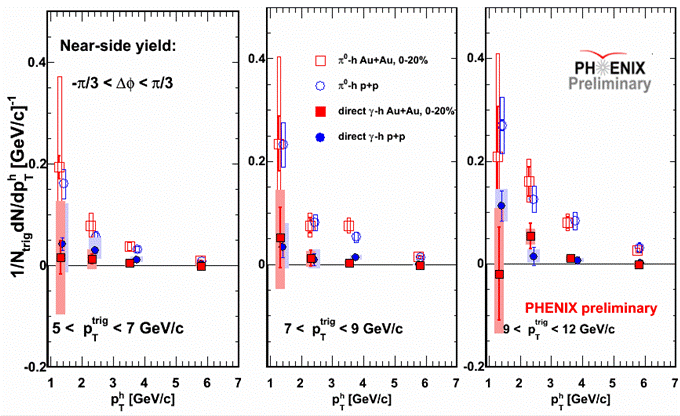}
\caption{(color online) Near side conditional yields for $\gamma_{dir}$-h and $\pi^0$-h correlations in
central Au+Au and p+p collisions from PHENIX~\protect\cite{megan_qm09}.}
\label{near_comp}
\end{figure}

Near side correlations, where the azimuthal angle between the 
$\gamma_{dir}$ and the hadron is small are very interesting in 
heavy ion collisions.  In hadron-hadron correlations, these
correlations come from same jet fragmentation.  In $\gamma_{dir}$-hadron
correlations in p+p collisions, these correlations exist only when the photon is from
jet fragmentation (for a discussion of fragmentation photons see Ref.~\cite{hanksqm09}).
Thus, the yield in p+p collisions of near side hadrons per trigger should be much smaller for
$\gamma_{dir}$-h than $\pi^0$-h correlations, as observed in Fig.~\ref{near_comp}.
The near side $\gamma_{dir}$-h correlations in Au+Au collisions are consistent with measurements
from p+p collisions. 

\begin{figure}[t]
\includegraphics[width=0.45\textwidth]{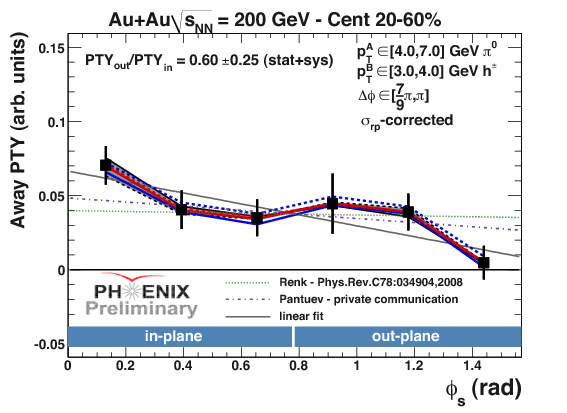}
\includegraphics[width=0.45\textwidth]{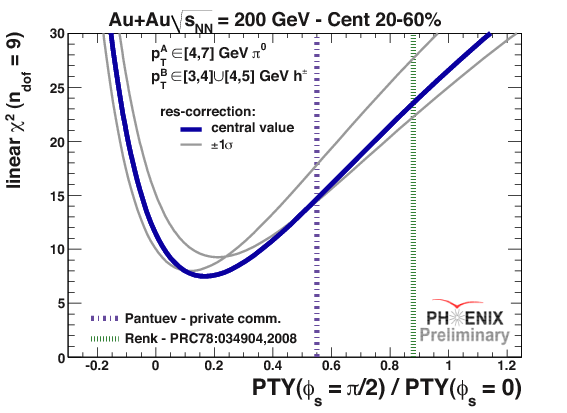}
\caption{(color online) Left: Away side conditional yield as a function of the angle of the
trigger with respect to the reaction plane for $\pi^0$-h correlations in mid-central
Au+Au collisions. Right: $\chi^2$ distribution for the ratio of the 
conditional yield out of plane to the conditional yield in plane 
using $4<p_{T,\pi^0}<7$GeV/c and hadrons with $3<p_{T,h}<4$GeV/c and
4$<p_{T,h}<5$GeV/c in mid-central Au+Au 
collisions~\protect\cite{mccumber_wwnd09,chen_qm09}.}
\label{rxnp_fig}
\end{figure}

Another sensitive measure of the path length dependence of the energy loss
is $\pi^0$-h correlations where the trigger is selected based on its angle
with respect to the reaction plane.  When the trigger is aligned with (out of) the reaction
plane the di-jet system sees a short (long) medium path length.  When the
energy loss through the core of the collision system is extremely large
(all di-jets are completely suppressed if they cross the center), then
the orientation with the trigger in the reaction plan will have a smaller
away side yield than when the trigger is out of plane; tangential di-jets 
which do not cross the center of the collision system will be favored.
If the di-jet suppression increases with the path length through the matter
however the suppression will be greater when the trigger is oriented out of the reaction
plane since that maximizes the away side path length.  Results from PHENIX
shown in Fig.~\ref{rxnp_fig}~\cite{mccumber_wwnd09}. show that the data favor the
latter case and the data favor a stronger reaction plane dependence than
expected from some theoretical models~\cite{pantuev,pantuev_jetp,renk_rxnp}.

\subsection{Heavy Flavor Conditional Yields}

It has been known for some time that electrons from the decay of heavy
mesons ($D$s and $B$s) are suppressed more than can typically be
accounted for in radiative energy loss calculations~\cite{ppg066,star_electron}.  There have been
a variety of theoretical attempts to explain this including collisional energy 
loss~\cite{coll_el}, recombination~\cite{sorensen_hq, ko_hq}, and in-medium
hadron formation~\cite{vitev_hq}.  The next experimental step is to correlate
the heavy quark electrons with other hadrons in the event in the same manner as is
done for $\pi^0$-h and $\gamma_{dir}$-h correlations.  

\begin{figure}
\includegraphics[width=0.55\textwidth]{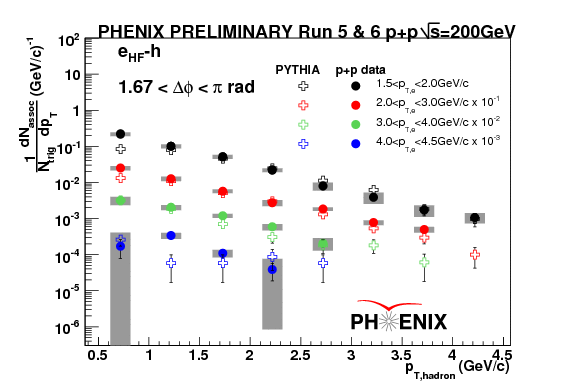}
\includegraphics[width=0.44\textwidth]{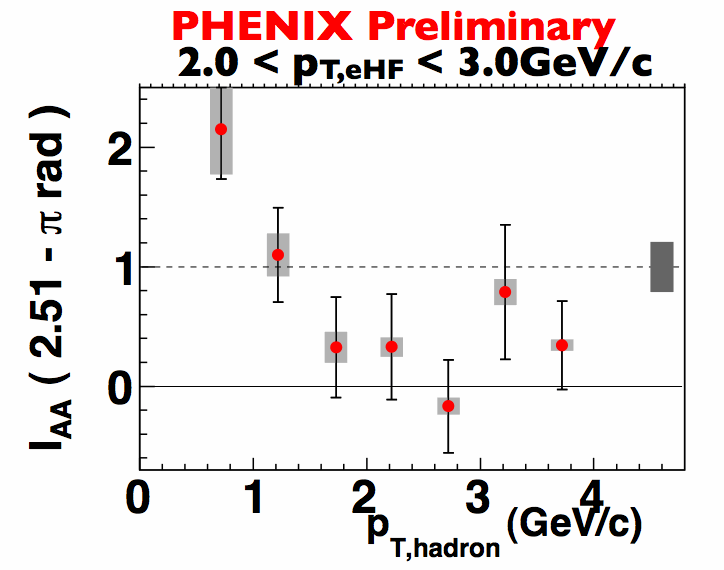}
\caption{(color online) Away side $e_{HF}$-h conditional yields in p+p collisions 
(left)~\protect\cite{sickles_wwnd09} and $I_{AA}$ (right).}
\label{eh}
\end{figure}

This measurement is severely complicated by the large
number of background electrons from $\pi^0$ Dalitz decays and photon conversions and
low statistics.
STAR has developed a method of tagging photonic electrons via the
invariant mass of $e^+e^-$ pairs~\cite{wang_qm08}. 
PHENIX has established a method analogous to that used for the $\gamma_{dir}$-h
correlations shown above to statistically subtract correlations from photonic 
sources~\cite{sickles_wwnd09}.  Figure~\ref{eh} shows results from p+p collisions compared to charm
production simulated in PYTHIA~\cite{pythia} (left)~\cite{sickles_wwnd09} 
and away side $I_{AA}$ which
suggests some suppression of away side hadrons opposite to electrons from heavy flavor
decay (right).  

Interpretation of these measurements is complicated because electrons from heavy flavor
decay come both from charm and bottom production.  Experimental efforts~\cite{ppg094,biritz_qm09} to 
determine the relative contribution of these sources in p+p collisions have agreed with 
Fixed Order Next to Leading Log calculations~\cite{fonll}.  However,
the relative contributions to the electron sample in heavy ion collisions could
be modified from p+p due to the effects of the medium.
In the $p_{T,e}$ range measured here the electrons are expected to mainly be from $D$ meson
decay.  Higher statistics measurements in the 2010 RHIC run might allow these measurements to be extended to higher
$p_T$ where bottom contributions are expected to be more significant.  
With detector upgrades capable of displaced vertex measurements it will be possible to 
distinguish charm and bottom triggered correlations.  Those measurements will be crucial 
in understanding the suppression of the single electron spectra in Au+Au collisions.

\subsection{Background Estimation and Two Source Model}
Jet correlation measurements in heavy ion collisions depend on a 
procedure to estimate and remove the combinatorial background
from pairs which do not come from the same di-jet production.  
Combinatorial pairs are assumed to only be angularly correlated through the reaction
plane and the $v_2$ values of the trigger and partner 
hadrons are assumed to be independent.  The shape of the combinatorial background in
$\Delta\phi$ is then, $b_0(1+2v_{2,trig}v_{2,part}\cos(2\Delta\phi))$.

$b_0$ is often estimated
by assuming the minimum in the jet function has zero yield~\cite{zyam}.
There are a number of issues associated with this assumption.  At the high
momenta considered here the fluctuation of the background level due to statistical
fluctuations in the correlation functions is a large source of uncertainty and
bias~\cite{bkgd_paper}.  Also in heavy ion collisions the jet induced correlations
are much wider than in p+p collisions making the assumption of a zero yield region
less desirable.
The use of the Absolute Background Subtraction method~\cite{bkgd_paper}
is preferable because of its stability in cases of low statistics and wide jets.
 Even when the combinatorial
background is small compared to the jet yield
 (as it is at high $p_T$) the uncertainty can be large compared to
the jet conditional yields.

\section{Conclusions and Outlook}
High $p_T$ identified spectra and correlation measurements give a significantly more nuanced picture
of parton-medium interactions than unidentified hadron measurements alone.  Contrary to 
expectations of high $p_T$ particle production by vacuum fragmentation alone, excess $p$ and
$\bar{p}$ in heavy ion collisions persist out to $p_T \approx$10GeV/c.  This could be evidence of
significant higher-twist baryon production and will be further investigated with higher statistics
data, correlations, and $v_2$ measurements.
In contrast both $\pi^0$-h and $\gamma_{dir}$-h correlations are consist with each other and show little 
dependence on the $p_T$ of either the trigger or associated particle.  
Initial measurements of $e_{HF}$-h correlation results have large statistical errors, but
show some evidence for away side suppression.  Future measurements with higher statistics and charm and bottom
separation will enable tomography of heavy flavor energy loss.  While existing single and di-hadron measurements
show that the created matter leads to a large energy loss, precision measurements with identified particles
will allow robust determination of the mechanism of energy loss as well as the mechanisms of hadron
formation.

\section{Acknowledgements}
I thank Andrew Adare, Bertrand Biritz, Megan Connors, Ahmed Hamed,
Mike McCumber, Bedenga Mohanty, Dave Morrison, Jamie Nagle, 
Paul Stankus and
Jiayin Sun interesting discussions and providing me with plots.
I thank the Quark Matter 2009 organizers for a very interesting and enjoyable conference.  This work is 
supported by the U. S. Department of Energy under contract DE-AC02-98CH1-886.

\bibliography{sickles_qm09_proceedings}

\begin{thebibliography}{10}
\expandafter\ifx\csname url\endcsname\relax
  \def\url#1{\texttt{#1}}\fi
\expandafter\ifx\csname urlprefix\endcsname\relax\def\urlprefix{URL }\fi

\bibitem{sevil_qm09}
S.~Salur, these proceedings, arXiv:0907.4536.

\bibitem{lai_qm09}
Y.~Lai, et~al., these proceedings, arXiv:0907.4725.

\bibitem{friesprc}
R.~J. Fries, et~al., Phys. Rev. C68 (2003) 044902.

\bibitem{hwa1}
R.~Hwa, C.~B. Yang, Phys. Rev. C70 (2004) 024904.

\bibitem{stananne}
S.~Brodsky, A.~Sickles, Phys. Lett. B668 (2008) 111--115.

\bibitem{vitev}
I.~Vitev, M.~Gyulassy, Phys. Rev. Lett 89 (2002) 232301.

\bibitem{ppg079}
A.~Adare, et~al., Phys. Rev. C77 (2008) 064907.

\bibitem{dss07}
D.~de~Florian, R.~Sassot, M.~Stratmann, Phys. Rev. D76 (2007) 074033.

\bibitem{reco_review}
R.~Fries, V.~Greco, P.~Sorensen, Ann. Rev. Nucl. Part. Sci. 58 (2008) 177--205.

\bibitem{conversion_ko}
W.~Liu, C.~Ko, B.~Zhang, Phys. Rev. C75 (2007) 051901.

\bibitem{conversion_fries}
W.~Liu, R.~Fries, Phys. Rev. C77 (2008) 054902.

\bibitem{shengli}
S.~Huang, et~al., J. Phys. G 36 (2009) 064061.

\bibitem{ppg072}
A.~Adare, et~al., Phys. Lett. B 649 (2007) 359--369.

\bibitem{yichunqm09}
Y.~Xu, et~al., these proceedings, arXiv:0907.4644.

\bibitem{stardijet}
C.~Adler, et~al., Phys. Rev. Lett. 97 (2006) 162301.

\bibitem{ppg083}
A.~Adare, et~al., Phys. Rev C78 (2008) 014901.

\bibitem{ppg032}
S.~S. Adler, et~al., Phys. Rev. Lett. 97 (2006) 052301.

\bibitem{starlowpt}
C.~Adler, et~al., Phys. Rev. Lett. 95 (2005) 152301.

\bibitem{frantzhp08}
J.~Frantz, et~al., arXiv:0901.1393.

\bibitem{ppg090}
A.~Adare, et~al., Phys. Rev. C 80 (2009) 024908.

\bibitem{megan_qm09}
M.~Connors, et~al., these proceedings, arXiv:0907.4571.

\bibitem{hamed_qm09}
A.~M. Hamed, et~al., these proceedings, arXiv:0907.4523.

\bibitem{ppg029}
S.~S. Adler, et~al., Phys. Rev. D74 (2006) 072002.

\bibitem{zoww_gamma}
H.~Zhang, et~al., Phys. Rev. Lett. 103 (2009) 032302.

\bibitem{zoww_hadron}
H.~Zhang, et~al., Phys. Rev. Lett. 98 (2007) 212301.

\bibitem{renk_gamma}
T.~Renk, Phys. Rev. 80 (2009) 014901.

\bibitem{renk_hadron}
T.~Renk, Phys. Rev. C78 (2008) 014903.

\bibitem{hanksqm09}
A.~Hanks, et~al., these proceedings, arxiv:0907.4825.

\bibitem{mccumber_wwnd09}
M.~McCumber, et~al., arXiv:0905.0429.

\bibitem{pantuev}
V.~Pantuev, nucl-ex/0610002.

\bibitem{pantuev_jetp}
V.~Pantuev, JETP Lett. 85 (2007) 104.

\bibitem{renk_rxnp}
T.~Renk, Phys. Rev. C78 (2008) 034904.

\bibitem{chen_qm09}
C.~Chen, et~al., these proceedings, arXiv:0907.4820.

\bibitem{ppg066}
A.~Adare, et~al., Phys. Rev. Lett. 98 (2006) 172301.

\bibitem{star_electron}
B.~Abelev, et~al., Phys. Rev. Lett. 98 (2007) 192301.

\bibitem{coll_el}
S.~Wicks, et~al., Nucl. Phys A784 (2007) 426--442.

\bibitem{sorensen_hq}
P.~R. Sorensen, X.~Dong, Phys. Rev. C74 (2006) 024902.

\bibitem{ko_hq}
Y.~Oh, et~al., Phys. Rev. C79 (2009) 044905.

\bibitem{vitev_hq}
A.~Adil, I.~Vitev, Phys. Lett. B649 (2007) 139,146.

\bibitem{wang_qm08}
G.~Wang, et~al., J Phys. G 35 (2008) 104107.

\bibitem{sickles_wwnd09}
A.~Sickles, et~al., arXiv:0905.2112.

\bibitem{pythia}
T.~Sjoestrand, et~al., Comp. Phys. Comm 135 (2001) 238.

\bibitem{ppg094}
A.~Adare, et~al., Phys. Rev. Lett. 103 (2009) 082002.

\bibitem{biritz_qm09}
B.~Biritz, et~al., these proceedings, arXiv:0907.3937.

\bibitem{fonll}
M.~Cacciari, P.~Nason, R.~Vogt, Phys. Rev. Lett. 95 (2005) 122001.

\bibitem{zyam}
N.~N. Ajitanand, et~al., Phys. Rev. C72 (2005) 011902.

\bibitem{bkgd_paper}
A.~Sickles, M.~McCumber, A.~Adare, submitted to Phys. Rev. C, arXiv:0907.4113.

\end{thebibliography}
\bibliographystyle{elsart-num}

\end{document}